# Towards Training Set Reduction for Bug Triage

Weiqin Zou   Yan Hu[*]
School of Software
Dalian University of Technology
Dalian, China
zou@mail.dlut.edu.cn   huyan@dlut.edu.cn

Jifeng Xuan
School of Mathematical Sciences
Dalian University of Technology
Dalian, China
xuan@mail.dlut.edu.cn

He Jiang[*]
School of Software
Dalian University of Technology
Dalian, China
jianghe@dlut.edu.cn

*Abstract*—**Bug triage is an important step in the process of bug fixing. The goal of bug triage is to assign a new-coming bug to the correct potential developer. The existing bug triage approaches are based on machine learning algorithms, which build classifiers from the training sets of bug reports. In practice, these approaches suffer from the large-scale and low-quality training sets. In this paper, we propose the training set reduction with both feature selection and instance selection techniques for bug triage. We combine feature selection with instance selection to improve the accuracy of bug triage. The feature selection algorithm $\chi^2$-test, instance selection algorithm Iterative Case Filter, and their combinations are studied in this paper. We evaluate the training set reduction on the bug data of Eclipse. For the training set, 70% words and 50% bug reports are removed after the training set reduction. The experimental results show that the new and small training sets can provide better accuracy than the original one.**

*Keywords-bug triage; training set reduction; feature selection; instance selection; software quality*

## I. INTRODUCTION

Bug fixing is a significant and time-consuming process in software maintenance [3]. For a large-scale software project, the number of daily bugs is so large that it is impossible to handle them without delaying [2]. The work of managing bugs increases the cost of software quality maintenance [17]. Many software projects use a bug tracking system to store and manage bugs submitted by *users*, including end users, testers, and developers [1]. The bug tracking system provides a platform, where users can communicate with each other during the bug fixing process. Bugzilla[1] is such a bug tracking system, which is used by many large open source software projects. Based on the bug tracking system, the developers can easily search and maintain all the existing bugs.

*Bug triage*, an important step for bug fixing, is to assign a new bug to a relevant developer for further handling [3]. A general method for bug triage is to assign bugs manually. In practice, due to the frequent changes of software development teams, it is difficult to identify the correct developer in manual triage [3]. Taking Eclipse[2] as an example, Anvik reports that an average of 37 bugs per day are submitted to the bug tracking system and 3 person-hours per day are required for the manual triage [2]; the empirical study by Jeong et al. shows that 44% of bugs have been assigned to the wrong developer after the first assignment [4]. To solve these problems, some machine learning algorithms are employed to conduct automatic bug triage [1], [3], [4], [5], [6]. Most of the bug triage approaches are based on text categorization [3]. However, these approaches suffer from two problems. On one hand, due to the large number of bugs, it is necessary to collect large-scale training sets of bugs to obtain good results for bug triage [3]. For example, in our experiments, over 9000 bug reports and 30000 words are employed to train a classifier for Eclipse. It may cost much time to directly use the large-scale training set in the bug triage process. On the other hand, the quality of the original bug reports is not good enough. Low-quality bug reports may mislead the triage approach to assign bugs to wrong developers [6], [16].

In this paper, we propose a training set reduction approach using the combination of feature selection and instance selection. Our training set reduction approach is a two-phase process, which applies feature selection (removing unnecessary words) before or after instance selection (removing unnecessary reports). We use the typical feature selection algorithm, $\chi^2$-test (CHI) [12] and instance selection algorithm, Iterative Case Filter (ICF) [14] to build a small and effective training set by removing the noisy and redundant words and bug reports. In the experiments, we evaluate our training set reduction on the Eclipse project. After the training set reduction, 70% words and 50% bug reports are removed. The results show that the experiments on the reduced training sets can obtain better accuracies than that on the original training set.

This paper makes the following contributions.
- We propose the training set reduction approach for bug triage by combining feature selection with instance selection. To our knowledge, this is the first work to improve the performance of bug triage by reducing the scale of the training set.
- We provide a comparative study on the effect of different orders of combinations. The results show that the effectiveness of the training set reduction can be influenced by changing the order of two phases.

The remainder paper is organized as follows. In Section II, we briefly introduce bug triage and give two motivating examples for training set reduction. Section III gives the details of our training set reduction. In Section IV, we present the experimental results and analysis. We give the threats to validity of our work in Section V. Section VI lists the related work. We briefly conclude this paper and present our future work in the final section.

---

[*] Corresponding Author
[1] Bugzilla, http://www.bugzilla.org/.
[2] Eclipse, http://www.eclipse.org/.

## II. BUG TRIAGE AND MOTIVATING EXAMPLES

In a bug tracking system, a software bug is stored as a *bug report*, which is the textual form for describing the details of a bug [3]. When a user submits a bug to the bug tracking system, a bug report is filled to provide information to help identifying and reproducing the bug. For example, a bug report may contain the items for recording the summary, the detailed description, the buggy product, the sub-component, and the operating system. Among these items, the summary and the detailed description are always used in bug triage since they can provide adequate information of bugs [3].

Čubranić & Murphy have proposed the first work of bug triage, in which the bug assignment is innovatively mapped to the text categorization [1]. To improve the accuracy of bug triage, a recommendation list is employed to predict multiple relevant developers [3]. Fig. 1 shows the process of the text categorization approach for bug triage.

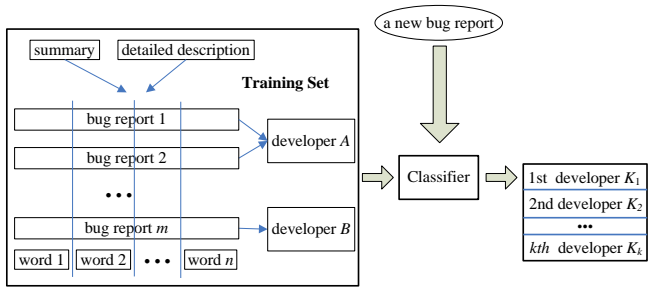

Figure 1. The text categorization approach for bug triage.

In Fig. 1, the training set is viewed as a *text matrix*. Each row of the matrix indicates one bug report while each column of the matrix indicates one word. However, due to the large scale and the low quality, some bug reports may bring noise and redundancy into the data set. In the following part of this section, we will give two examples of bug reports to present the need for the training set reduction.

We take the bug report with ID 205900 in Eclipse as an example to study the words of bug reports. This bug report describes a broken error of the discovery repository.

Bug report 205900: ....*[Plug-ins]* all **installed** correctly and do not **show** any errors in Plugin configuration view. Whenever I try to add a *[diagram name]* diagram, the wizard can not be **started** due to a **missing** *[class name] class...*

In this bug report, some words, e.g., *installed*, *show*, *started*, and *missing*, are commonly used for describing bugs. For a text categorization based approach, these common words are not helpful for predication. Thus, we want to remove these words to improve bug triage (actually, these four words are removed in the experiments in Section IV after the training set reduction). However, for the text categorization, these redundancy words cannot be removed directly. Thus, a relevant technology for bug triage needs to be developed.

We take two bug reports with ID 200019 and ID 204653 as examples to study the relationship between bug reports (the detailed descripitons are omitted).

Bug report 200019: *Argument popup not highlighting the correct argument...*
Bug report 204653: *Argument highlighting incorrect...*

These two bug reports are duplicate (*duplicate*，denotes that more than one bug reports describe one software fault). The textual forms of these two bug reports are similar. Thus, a bug report may be chosen as the representative one of these bug reports. Thus, we want to use some technology to remove one of these bug reports (actually, the bug report with ID 200019 is removed while the bug report with ID 204653 is kept in the experiments in Section IV). In this paper, we focus on the technology to remove the extra words and bug reports for bug triage.

## III. TRAINING SET REDUCTION FOR BUG TRIAGE

In this section, we present our training set reduction approach for bug triage. In our work, a feature selection algorithm and an instance selection algorithm are combined to reduce noisy or redundant information in the training set of bug triage.

### A. Training Set Reduction

Motivated by the examples in Section II, we employ both feature selection and instance selection algorithms to remove the unnecessary features and instances. Drawn on the experience in text categorization, a feature in bug triage indicates a word while an instance indicates a bug report. Therefore, the goal of our work is to reduce the text matrix of the training set on two dimensions, i.e., the word dimension and the bug report dimension.

To generate a reduced training set on two dimensions, we employ a two-phase combination to take advantage of both feature selection and instance selection (a phase is a process applying one algorithm). The reduced training set is applied to replace the original training set of bug triage. To distinguish the combinations, we investigate the order of the two phases. Given a feature selection algorithm $FS$ and an instance selection algorithm $IS$, we use $FS \rightarrow IS$ to denote first applying $FS$ and then $IS$. On the other hand, $IS \rightarrow FS$ denotes first applying $IS$ and then $FS$.

Based on the definition of feature selection [12], the words with small objective values are removed to improve the prediction performance. Thus, the high-quality words are left to train the classifier. As an input parameter, a threshold of final number of words is used to terminate the process of feature selection. Different from a feature selection algorithm, the terminal condition of an instance selection algorithm is usually based on a heuristic rule, which can be set according to the applications [14]. In this paper, we set the final number of bug reports as the terminal condition.

In Algorithm 1, we briefly show the training set reduction with the order $FS \rightarrow IS$. As described in the Algorithm 1, we first apply feature selection, and then apply instance selection. Before feature selection, we calculate objective values for all the words. After feature selection on the original data set, a new and small data set with the representative words is generated. Then we perform instance selection on the new data set. After the training set reduction, the final training set is generated. Note that in Step 2), some bug reports may be empty during feature selection (i.e., all the words of a bug report are removed). Such zero-word bug reports can be removed in the instance selection algorithm in Step 3).

---
**Algorithm 1**. Training set reduction with $FS \rightarrow IS$
---
**Input**: training set $T$ with $n$ words and $m$ bug reports,
  feature selection algorithm $FS$, final number of words $n_F$,
  instance selection algorithm $IS$, final number of bug reports $m_I$,
  order of the combination $FS \rightarrow IS$
**Output**: reduced training set $T_{FI}$ for bug triage

1) Apply $FS$ to $n$ words of $T$ and calculate objective values for all the words;
2) Select the top $n_F$ words of $T$ and generate a training set $T_F$;
3) Apply $IS$ to $m$ bug reports of $T_F$;
4) Terminate $IS$ when the number of bug reports is equal or less than $m_I$ and generate the final training set $T_{FI}$.
---

For the combinations of $FS$ and $IS$, we will present the results of $FS \rightarrow IS$ and $IS \rightarrow FS$, respectively. All the combination results are shown in details in the experiment section. In the following parts of this section, we briefly introduce the feature selection algorithm CHI and the instance selection algorithm ICF used in our work.

### B. CHI Feature Selection Algorithm

Feature selection is a standard technology to reduce the features of large-scale data sets in machine learning. Since many feature selection algorithms have been investigated for text categorization, we select a typical algorithm in our work, i.e., $\chi^2$-test (CHI)[3]. Yang & Pedersen [12] have given a comparative study on five feature selection algorithms and have reported that CHI can outperform the other algorithms in the study.

CHI is a typical feature selection algorithm, which measures the dependence between words and developers [12]. Among the bug triage approaches, the vocabulary-based expertise model by Matter et al. [5] also focuses on the relationship between words and developers. For each developer $d_i$ in the developer set $D$, given a word $w$, let $A$ be the co-occur times of $w$ and $d_i$, $B$ be the times of $w$ occurring without $d_i$, $C$ be the times of $d_i$ occurring without $w$, and $D$ be the times of neither $w$ nor $d_i$ occurring. In practice, the maximum value or the average value of each developer can be used to evaluate the given word. In this paper, we use the maximum value of the dependence as (2).

$$CHI(w) = \max_{d_i \in D}(\chi^2(w,d_i)) = \max_{d_i \in D} \frac{m \times (A \times D - C \times B)}{(A+C) \times (B+D) \times (A+B) \times (C+D)} \quad (2)$$

---
[3] CHI tool from NEU-NLP lab, http://www.nlplab.com/.

### C. ICF Instance Selection Algorithm

Besides feature selection, instance selection is a technology to reduce the number of instances and to enhance the training set quality. According to [13], we select Iterative Case Filter (ICF) [14] as the instance selection algorithm in our work.

ICF is an instance selection algorithm based on the k-Nearest Neighbor algorithm (kNN) [7]. The process of ICF consists of two steps, namely noise filtering and instance condensing. During the iteration of ICF, kNN is used to evaluate whether an instance is representative for the classes.

## IV. EXPERIMENTAL RESULTS AND ANALYSIS

In this section, we present the experiments of the training set reduction for bug triage. First, we show the data preparation for applying the training set reduction. Then, we give the experimental setup. Next, we show the accuracy rates of each single algorithm. Finally, we detailed present the results and the analysis for the experiments of Eclipse.

All the experiments are conducted on *Debian* under the platform of a PC with Intel Core 2.8 GHz CPU and 4 GB memory. Besides using the open access tools, we implement the ICF algorithm and the Naive Bayes classifier in our work.

### A. Data Preparation

To evaluate the experimental results of the training set reduction, we employ the bug data of Eclipse. We choose Eclipse since the training set is easy to obtain and the labeling heuristic of bug triage can work well on it [3].

We select a set of continuous bug reports as a data set for the experiments. According to the classic study of bug triage [3], during the three months before and after a project release, the submission of bugs is very active. Thus, the data set in our experiments is chosen around the project release. We choose the bug reports with IDs from 200001 to 220000. To avoid the influence of the out-of-date bugs, we remove the duplicate bugs at the early stage and retain the ones around the project release.

To build the data set in the experiments, we follow the existing work (e.g., [3], [6]) to select the fixed bugs and duplicate bugs. As a result, 12385 bug reports are left. To get a more effective training set, we remove the inactive developers, who have fixed less than 10 bugs since no sufficient information is offered to predict the bug-fixing capabilities of such developers. After the above steps, the final number of bug reports for Eclipse is 11313. Table I shows the detailed information of the final data set.

Before training a classifier for the bug reports, each bug report should be labeled to identify the correct developer [1]. We label the correct developer according to the labeling heuristic [3]. For a new-coming bug report, the summary and the detailed description are the most representative items, which are used for manual bug triage [1]. Thus, for each bug report, we use the summary and the detailed description (the first long description in Bugzilla) to obtain the words. Then we use the BoW toolkit's [15] built-in text categorization techniques (calculating the term frequency and removing the stopwords) to convert each bug report into a word vector based on the vector space model [3]. We do not use the stemming technique since an emperical study has indicated that the stemming technique is not helpful to bug triage [1].

TABLE I. THE DATA SET OF THE OPEN SOURCE PROJECT ECLIPSE

| Item | Value |
| --- | --- |
| Number of instances | 11313 |
| Number of features | 35724 |
| Number of developers | 267 |
| Minimum number of bugs fixed per developer | 10 |
| Maximum number of bugs fixed per developer | 284 |

### B. Experimental Setup

To compare the results of our experiments, we employ the Naive Bayes classifier as the bug triage approach [1], [3], [4], [6]. The kernel idea of Naive Bayes is to calculate the posterior probability for each new-coming bug report from the prior probability and the likelihood of the existing bug reports [1]. In our experiments, the Naive Bayes approach without the training set reduction is employed as a baseline (called "Origin" in the experiments).

To train a classifier for bug triage, we split the data set into a training set and a test set. The 5-fold cross-validation is used to calculate the average of the results to avoid over-fitting problem [7]. To improve the quality of bug triage, we follow the existing work ([3], [4], [5], [6]) to use a recommendation list. A list with the size $k$ can provide $k$ developers as the prediction result for each new-coming bug report.

As the evaluation criteria, we employ the accuracy rate, the precision rate, and the recall rate to report the experimental results. We give the definition of the evaluation criteria as follows ($k$ denotes the size of the recommendation list).

$$Accuracy_k = \frac{\text{\# correct relevant developers}}{\text{\# all bug reports in test set}} \quad (3)$$

$$Precision_k = \frac{\text{\# correct relevant developers}}{\text{\# relevant developrers} \times k} \quad (4)$$

$$Recall_k = \frac{\text{\# correct relevant developers}}{\text{\# correct developers}} \quad (5)$$

The accuracy rate is the most significant evaluation criterion for bug triage since it measures the quality of prediction [3], [4], [5], [6]. The other two criteria, the precision rate and the recall rate, are used to measure the relevance and correctness of bug triage [3], [5].

### C. Results of Feature Selection and Instance Selection

Before given the results of the training set reduction, we show the results of each single algorithm in this part.

Fig. 2 presents the accuracy rates of the CHI and ICF on Eclipse. For CHI, we select 10%, 30%, and 50% as the ratio of the final number of words, respectively. Such parameter setup is based on the experience of the text feature selection study [12]. For ICF, we set the ratio of the final number of bug reports as 30%, 50%, and 70%, respectively. The ratio value set up is based on the study of text instance selection [13]. Due to the algorithm mechanism of ICF, it is possible that different ratios can lead to the same results (i.e. the same instances are selected). In our experiments with ICF, we obtain the identical result for the ratio of 30%, 50%, and 70%. Thus, we only draw one curve for ICF in Fig. 2.

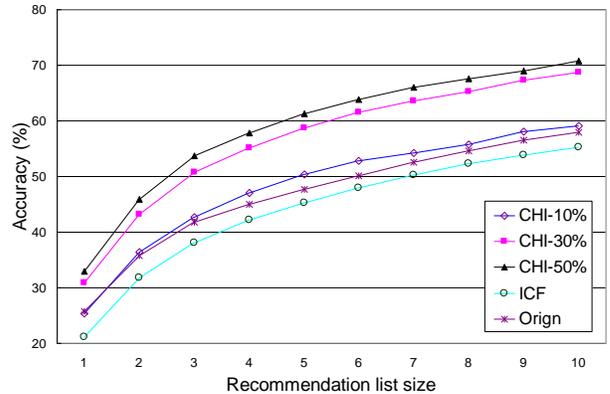

Figure 2. The accuracy rates of CHI and ICF on Eclipse

From Fig. 2, it can be found that CHI works much better than the original experiment. When the recommendation list size is one, the improvement of the accuracy rate for Eclipse is up to 13%. Meanwhile, we can see that CHI achieves good performance if 30% or 50% is selected as the ratio of the final number of words. In contrast with the results of feature selection, the accuracy rate of instance selection is not better than the original result.

Although the accuracy rate of instance selection is not ideal for bug triage, we find that the number of instances is significantly reduced by the instance selection algorithm. For bug triage, if we can improve the accuracy rate of instance selection, ICF is useful for reducing the training sets.

### D. Results of Training Set Reduction

Based on the results of feature selection and instance selection, we can see that both feature selection and instance selection can reduce the scale of training set (for words or for bug reports) and feature selection can improve the accuracy rates. These results provide the basis for our combined training set reduction. In other words, we use the accuracy improvement obtained by feature selection to cover the loss of instance selection.

To terminate the process of the training set reduction, each algorithm needs an input parameter of the final number of the words or bug reports, a general method to determine these parameter values is *parameter tuning* [12], [14]. Due to the time cost of parameter tuning, we only artificially allocate the parameter values in our experiments. For each algorithm, we choose the middle value of the final numbers of features or instances, i.e., we set 30% as the ratio of the final number of words and 50% as the ratio of the final number of bug reports. We present the results of the accuracy rates of the training set reduction on Eclipse in Table II. In the table, the recommendation list size is from 1 to 10. We show the results of the original experiment, the two single algorithms, and the two combinations.

TABLE II. ACCURACY RATES OF TRAINING SET ON ECLIPSE

| List size | Origin | CHI | ICF | CHI→ICF | ICF→CHI |
|---|---|---|---|---|---|
| 1 | 25.83 | 30.91 | 21.18 | 25.22 | **27.23** |
| 2 | 35.71 | 43.21 | 31.85 | 38.18 | **40.15** |
| 3 | 41.76 | 50.80 | 38.09 | 46.73 | **48.57** |
| 4 | 45.02 | 55.16 | 42.17 | 51.61 | **53.45** |
| 5 | 47.68 | 58.66 | 45.29 | 55.82 | **57.29** |
| 6 | 50.15 | 61.58 | 47.96 | 58.93 | **60.09** |
| 7 | 52.58 | 63.54 | 50.22 | 60.75 | **62.00** |
| 8 | 54.58 | 65.22 | 52.31 | 62.88 | **63.92** |
| 9 | 56.53 | 67.35 | 53.91 | 64.85 | **65.64** |
| 10 | 57.92 | 68.66 | 55.23 | 65.90 | **66.95** |

From the combination results, we can find that both CHI→ICF and ICF→CHI can provide better accuracies than the original experiment. Note that the single algorithm CHI can obtain some better accuracies than the combinations, e.g., on Eclipse with the recommendation list size one, the accuracy rate of ICF→CHI is 27.23% while that of CHI is 30.91%. However, the goal of our work is to reduce the scale of training set. Since ICF→CHI removes 50% bug reports more than CHI, the training set reduction of ICF→CHI is more effective. From Table II, we also can find out that CHI→ICF outperforms ICF→CHI, one reason for this fact is that some features may be removed with the instances during instance selection while very few instances are removed during feature selection [14].

Besides the accuracy rate, we employ the precision and recall rates to evaluate the performance of the training set reduction. We present the results in Fig. 3. It can be found that ICF→CHI provides the best precision and recall rates. CHI→ICF and CHI can generate better results than the other two approaches. Thus, to achieve the high precision and recall rates, ICF→CHI is an ideal decision for the training set reduction.

## V. THREATS TO VALIDITY

There are two threats of validity for our work. First, the conclusions from the experimental results cannot be directly transferred to other projects. In this paper, our experiments are based on parts of the bug data from the Eclipse and it is hard to model the differences from Eclipse to other projects. As a result, the experimental results on some other projects may be different from ours. Since most steps of our experiments are automatic, our experiments of the training set reduction can be adapted to other projects. To solve this problem, an empirical study on more projects may provide a detailed comparison.

Second, the precision rate in our work is very low. The reason for this fact is that the calculation rule is an adaptation from the precision of the recommendation system [5]. We only have one relevant developer for each new-coming bug report while many recommendation systems provide multiple classes for each instance. Therefore, the value of the precision rate is low in our experiments.

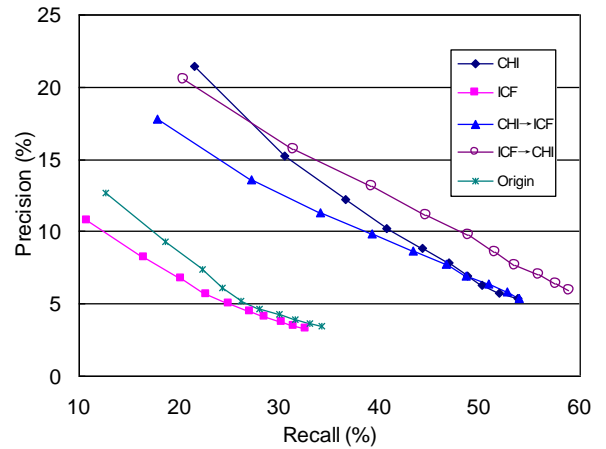

Figure 3. The precision and recall rates on the combinations of CHI and ICF on the data sets of Eclipse. The points of each curve denote the recommendation list size from one (left) to ten (right).

## VI. RELATED WORK

As to our knowledge, there is no study on feature selection and instance selection for the bug triage problem. The existing bug triage approaches are based on the text categorization. The first work of bug triage, proposed in 2004, is a supervised text categorization approach using Naive Bayes [1]. Anvik et al. [2], [3] extend this work with some other supervised learning algorithms (e.g., support vector machine), a recommendation list, and a complex labeling heuristic. To assist bug triage with the developer networks, Jeong et al. [4] propose a tossing graph approach to improve the accuracy of recommendation; Bhattacharya & Neamtiu [17] extend the tossing graph approach with the multi-feature technology and a fine-grained incremental learning; Matter et al. [5] use the vocabulary based expertise model to improve bug triage. To solve the problem of low quality of bug reports, Xuan et al. [6] propose a semi-supervised learning approach with a weighted recommendation list for bug triage.

When facing the new-coming bugs, removing the duplicate ones is a labor-intensive task for large-scale projects. Runeson et al. [9] propose an information retrieval approach to detect the duplicate bugs; Wang et al. [10] extend this approach with execution information. In contrast to removing duplicate bugs, Bettenburg et al. [8] consider adding the duplicate information to improve bug triage.

The most relevant work of this paper is the feature selection technology for bug prediction proposed by Shivaji et al. [11]. In this work, they employ the information gain algorithm to improve the quality of bug prediction. In this paper, we focus on the bug triage problem. Besides feature selection, we study the results of instance selection and combine feature selection with instance selection to reduce the scale of the training set. The goal of our work is to remove the noisy and redundant bug reports. To our knowledge, this is the first work of both feature selection and instance selection in software engineering.

## VII. Conclusion and Future Work

This paper is the first work of combining feature selection with instance selection to reduce the training set for the bug triage problem. The motivation of this work is to reduce the large scale of the training set and to remove the noisy and redundant bug reports for bug triage. Based on our setup, 70% of words and 50% of bug reports are removed. The experimental results show that the combinations of CHI and ICF can achieve better accuracy rates than that without the training set reduction. The results also indicate that the combination, ICF→CHI, is a good choice for the training set reduction.

In the future work, we plan to propose a unified approach to merge the tasks of feature selection and instance selection. In this paper, we focus on the combinations of the existing algorithms for the training set reduction. Since each algorithm in the combination is limited by the other one, it is necessary to develop a unified approach to integrate feature selection and instance selection.

Another future work is to apply the training set reduction of bug triage to other tasks to improve the software quality. Since machine learning becomes one of the powerful tools in software engineering, the training set reduction can be useful for the work based on machine learning.

## Acknowledgment

Many thanks to Dr. John Anvik with Department of Computer Science, University of Victoria for sharing the labeling heuristic for bug triage. Thanks the NEU-NLP lab for sharing the CHI tool to do feature selection. Thanks Runqing Wang with Dalian University of Technology for the implementation of a part of ICF algorithm. Thanks Zhilei Ren with Dalian University of Technology for his kind-hearted suggestions.

Our work is partially supported by the Natural Science Foundation of China under Grant No. 60805024 and the National Research Foundation for the Doctoral Program of Higher Education of China under Grant No. 20070141020.